\documentclass[letterpaper, 10 pt, conference]{ieeeconf}  

\IEEEoverridecommandlockouts                              

\usepackage{graphicx} 
\usepackage{amsmath} 
\usepackage{amsfonts}
\usepackage{mathtools} 
\usepackage{amssymb}
\usepackage{xcolor}
\usepackage{todonotes}
\usepackage{caption}
\usepackage{subcaption}
\usepackage{svg}
\usepackage{bm} 
\usepackage{mathrsfs}  
\usepackage{url}

\newcommand{\mf}[1]{{\mathbf{#1}}}

\newcommand{\vf}[1]{{\bm{#1}}}

\newcommand{\reviewquestion}[2]{#2}

\DeclareMathOperator*{\argmin}{arg\,min}

\makeatletter
\let\NAT@parse\undefined
\makeatother
\usepackage{hyperref}
\hypersetup{
    colorlinks=true,
    linkcolor=blue,
    filecolor=magenta,      
    urlcolor=cyan,
}
\begin{document}

\title{\reviewquestion{R2-Q6}{Velocity Level Approximation of Pressure Field Contact Patches}}

%
%
\author{Joseph Masterjohn, Damrong Guoy, John Shepherd, Alejandro Castro
\thanks{All the authors are with Toyota Research Institute, USA, {\tt\small firstname.lastname@tri.global}.}}

\maketitle

\begin{abstract}
  \reviewquestion{R2-Q1/R2-Q2/R3-Q1}{Pressure Field Contact (PFC) was recently
introduced as a method for detailed modeling of contact interface regions at
rates much faster than elasticity-theory models, while at the same time
predicting essential trends and capturing rich contact behavior. The PFC model
was designed to work in conjunction with error-controlled integration at the
acceleration level. Therefore a vast majority of existent multibody codes using
solvers at the velocity level cannot incorporate PFC in its original form. In
this work we introduce a discrete in time approximation of PFC making it
suitable for use with existent velocity-level time steppers and enabling execution
at real-time rates. We evaluate the accuracy and performance gains of our
approach and demonstrate its effectiveness in simulating relevant
manipulation tasks. The method is available in open source as part of Drake's
Hydroelastic Contact model.}

\end{abstract}

\begin{keywords}
  Contact Modeling, Simulation and Animation, Grasping, Dynamics.
\end{keywords}

%
\IEEEpeerreviewmaketitle

\section{Introduction}

\begin{figure}[!ht]
	\centering
	\includegraphics[height=1.55in]{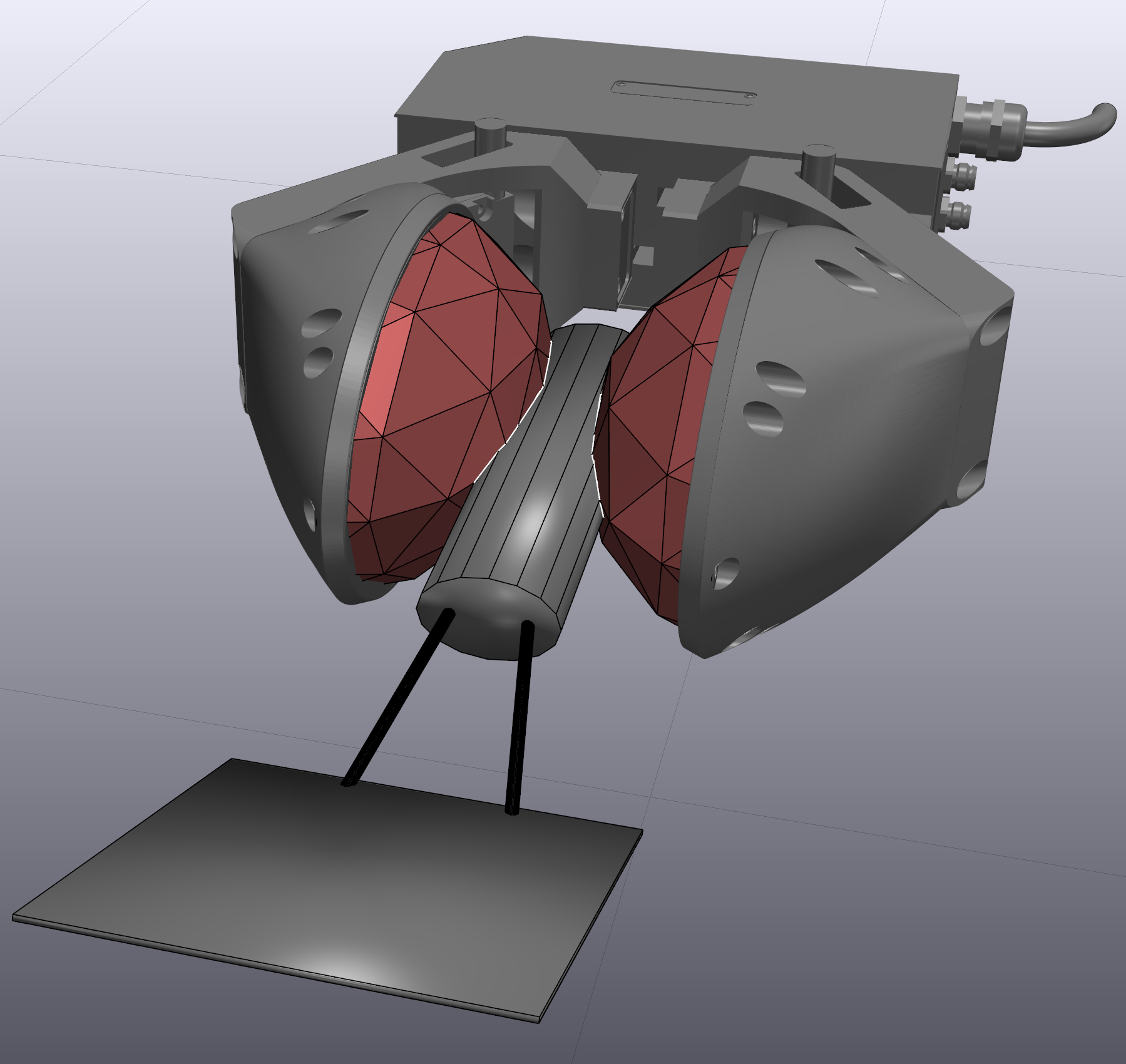}
    \includegraphics[height=1.55in]{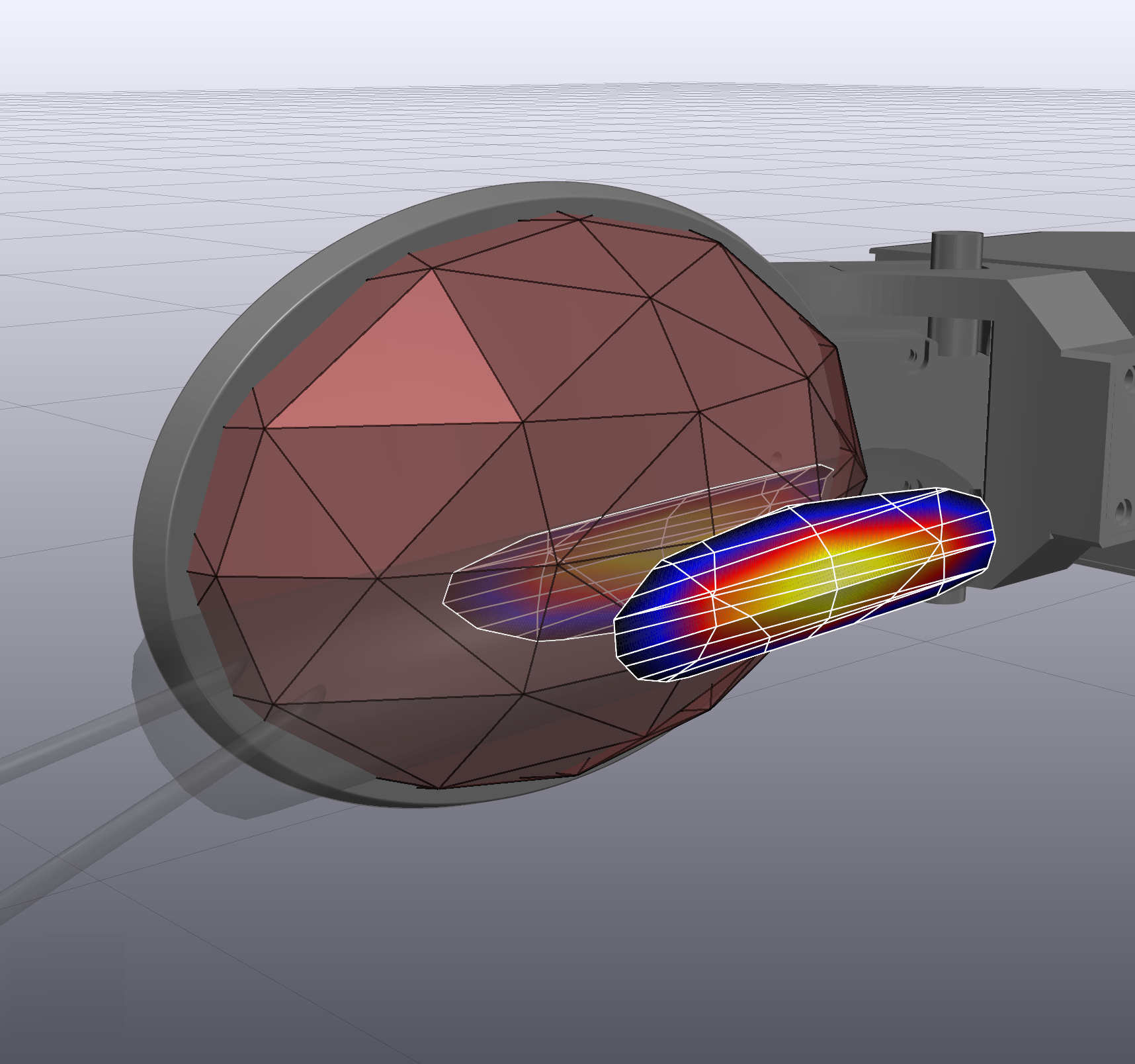}
	\caption{\label{fig:slip_control}
    Left: Contact geometry modeling the highly compliant \textit{Soft-bubble}
    gripper \cite{bib:kuppuswamy2020soft} holding a spatula. Right: One of the
    fingers is hidden to better show the simulated contact patch.
    \reviewquestion{R2-Q12}{With a coarse discretization,} the model is able to
    predict patch shape, size, and contact pressure (shown as colored contours)
    along with net forces and moments. Our polygonal tessellation of the contact
    surface combined with our velocity-level time stepping approximation enables
    this simulation to run at real-time rates.}
\end{figure}

\PARstart{T}{here} is a need for smooth, rich, artifact-free models of contact
between arbitrary geometries as encountered in modern robotics applications such
as grasping and manipulation, assistive and rehabilitative robotics,
prosthetics, and unstructured environments. Most often these applications
involve compliant surfaces such as padded grippers, deformable manipuland
objects or soft surfaces for safe human-robot interaction. Moreover, with the
emerging field of soft robotics, designers have begun to incorporate significant
compliance in their robot designs; consider for instance the
\textit{Soft-bubble} gripper \cite{bib:kuppuswamy2020soft} in Fig.
\ref{fig:slip_control} for which the accurate prediction of contact patches is
critical for meaningful sim-to-real transfer. Still, the rigid-body
approximation of contact is at the core of many simulation engines enabling them
to run at interactive rates.

Point contact is a useful and popular approximation of non-conforming contact
(e.g. contact between a sphere and a half-space), but it does not extend well to
conforming surfaces nor non-convex shapes. Localized compliance can be
incorporated using spring-dampers \cite{bib:catto_softconstraints}, Hertz theory
\cite{bib:luo2006compliant} and volumetric models \cite{bib:gonthier2007contact,
bib:wakisaka2017loosely}. However, while point contact modeling approaches are
fast, they are non-smooth, and extensions to arbitrary geometry often involve
non-physical heuristics \cite{bib:bullet_siggraph2015, bib:moravanszky2004fast}
that heavily influence the correctness and accuracy of simulation results
\cite{bib:erleben2018methodology}.

The Elastic Foundation Model \cite{bib:johnson1987contact} (EFM) computes rich
contact patches providing an alternative to point contact that can solve many of
its issues. However, current implementations \cite{bib:sherman2011simbody} need
highly refined meshes and can even miss contact interactions if coarse meshes
are used. The work in \cite{bib:elandt2019pressure} introduces \emph{pressure
field contact}; a modern rendition of EFM designed to work with coarse meshes at
a computational cost suitable for real-time simulation.
\reviewquestion{R3-Q14/R3-Q18}{While previous work focuses on smooth geometric
queries and continuous penalty forces, see for instance
\cite{bib:macklin2020local, bib:tang2012continuous, bib:geilinger2020add}, the
work in \cite{bib:elandt2019pressure} is different in that it introduces a
\emph{new contact model} rather than algorithms for an already existing
model.} An implementation of pressure field contact is available in open source
as part of Drake's \cite{bib:drake} Hydroelastic Contact model.
\reviewquestion{R2-Q5}{The implementation in Drake includes support for
primitive geometries such as spheres and boxes, convex meshes, rigid objects and
both triangular and polygonal tessellations, see Section
\ref{sec:hydro_overview} for details.} The hydroelastic contact model provides
rich information such as contact patch shape and pressure distribution, see Fig.
\ref{fig:slip_control} for an example.

\reviewquestion{R1-Q1/R2-Q2/R2-Q8/R3-Q1}{The hydroelastic contact model is
originally formulated at the acceleration level in \cite{bib:elandt2019pressure}
resulting in a system of ODEs advanced forward in time using error-controlled
integration. While error-controlled integration guarantees the accuracy of the
solutions, it comes with the additional cost of needing to compute error estimates and
taking smaller time steps during stick/slip transitions.

In contrast, popular simulation engines such as ODE \cite{bib:ode}, Dart
\cite{bib:dart}, Vortex \cite{bib:vortex}, MuJoCo \cite{bib:mujoco} and Drake
\cite{bib:drake} provide formulations at the velocity-level. In this approach,
time is advanced at discrete intervals of fixed size; contact impulses and the
resulting velocities are found by solving a challenging Nonlinear
Complementarity Problem (NCP), or some approximation of an NCP.

Our main contribution with this work is a discrete in time approximation of the
hydroelastic contact model that enables its use within existent simulation
engines formulated at the velocity-level. We derive an algebraic expression for
the rate of change of the pressure field in terms of local quantities and use it
to write an implicit in time approximation of the pressure field at the
centroids of mesh elements. Finally, we cast the problem in terms of an
equivalent set of compliant point contact forces that can be incorporated into
existent velocity-level formulations.}

This work also introduces a novel polygonal representation of the contact
surfaces introduced in \cite{bib:elandt2019pressure}. We strive to enable
simulation of contact rich patches, eliminate artifacts introduced by point
contact, and capture area dependent phenomena otherwise missed by point contact
while still performing at real-time rates. This is achieved with a complete
implementation in Drake \cite{bib:drake}.

\section{Multibody Dynamics With Frictional Contact}
\label{sec:discrete_time_stepping}

Here we closely follow the notation in our previous work
\cite{bib:castro2020transition, bib:castro2021sap} for consistency. However, we
point out that velocity-level engines with the capability to model compliant
point contact can incorporate the approximations introduced in this work to
model compliant contact patches using the hydroelastic contact model. For
stability, our approximations are implicit in time.

The state of our system is described by the generalized positions
$\mf{q}\in\mathbb{R}^{n_q}$ and the generalized velocities
$\mf{v}\in\mathbb{R}^{n_v}$, where $n_q$ and $n_v$ denote the number of
generalized positions and velocities, respectively. Time derivatives of the
configurations are related to the generalized velocities by
$\dot{\mf{q}}=\mf{N}(\mf{q})\mf{v}$, with
$\mf{N}(\mf{q})\in\mathbb{R}^{n_q\times n_v}$ the kinematic map.

\subsection{Contact Kinematics}
\label{sec:point_contact_approximation}

Given a configuration $\mf{q}$ of the system, we assume our geometry engine
reports a set $\mathscr{C}(\mf{q})$ of $n_c$ potential contacts between pairs of
bodies. The $i\text{-th}$ \emph{contact pair} in $\mathscr{C}(\mf{q})$ is
characterized by its location, a contact normal $\hat{\vf{n}}_i$, and the
\emph{signed distance} $\phi_i\in\mathbb{R}$, defined negative for overlapping
bodies. The relative velocity between the pair of bodies at the contact point is
denoted with $\vf{v}_{c,i}\in\mathbb{R}^3$. The normal and tangential components
of $\vf{v}_{c,i}$ are given by $v_{n,i} = \hat{\vf{n}}_i\cdot\vf{v}_{c,i}$ and
$\vf{v}_{t,i} = \vf{v}_{c,i}-v_{n,i}\hat{\vf{n}}_i$ respectively, so that
$\vf{v}_{c,i}=[\vf{v}_{t,i}\,v_{n,i}]$. We form vector
$\mf{v}_c\in\mathbb{R}^{3n_c}$ (bold, no italics) by stacking the velocities
$\vf{v}_{c,i}$. Contact velocities are related to generalized velocities by
$\mf{v}_c = \mf{J}(\mf{q})\mf{v}$, where $\mf{J}$ is the \emph{contact
Jacobian}.

\subsection{Contact Modeling}
\label{sec:contact_modeling}
A popular point contact model of compliance introduces a spring/damper at each
contact point to model the normal force $f_n$ as
\begin{equation}
    f_n = (-k\phi - d\,v_n)_+
    \label{eq:linear_normal_force},
\end{equation}
where $k>0$ is the point contact stiffness and $d>0$ is a coefficient of linear
dissipation, and $x_+ = \max(x, 0)$ is the positive part operator. Since we take
the positive part, the force is always repulsive. \reviewquestion{R3-26}{This
model can be cast as the equivalent complementarity condition
\cite{bib:lacoursiere2011spook}}
\begin{equation}
    0 \le \phi + d\,c\,v_n + c\,f_n \perp f_n \ge 0,
    \label{eq:nonpenetration_complementarity}
\end{equation}
where $c=k^{-1}$ is the compliance and $0 \le a\perp b \ge 0$ denotes
complementarity, i.e. $a \ge 0$, $b \ge 0$ and $a\,b=0$. Using the first order
approximation $\phi=\phi_0 + \delta t\,v_n$ where $\phi_0$ is the signed
distance function at the previous time step and $\delta t$ is step size, Eq.
(\ref{eq:nonpenetration_complementarity}) becomes a linear complementarity
condition between the velocities of the system and the contact forces. We use
the naught subscript to denote quantities evaluated at the previous time step
while no subscript is used for quantities evaluated at the next time step.
\begin{equation}
    0 \le \phi_0 + (\delta t+d\,c)v_n + c\,f_n \perp f_n \ge 0.
    \label{eq:complementarity_condition}
\end{equation}

The tangential component $\vf{f}_t\in\mathbb{R}^2$ of the contact forces is
modeled according to Coulomb's law of dry friction, which can be compactly
written as
\begin{equation}
    \vf{f}_t=\argmin_{\Vert\vf{f}\Vert\leq\mu f_{n}}\vf{v}_{t}\cdot\vf{f}
    \label{eq:maximum_dissipation_principle}
\end{equation}
where $\mu > 0$ is the coefficient of friction. Equation
(\ref{eq:maximum_dissipation_principle}) describes the \emph{maximum dissipation
principle}, which states that friction forces maximize the rate of energy
dissipation. In other words, friction forces oppose the sliding velocity
direction. Moreover, Eq. (\ref{eq:maximum_dissipation_principle}) states that
contact forces $\vf{f}_c$ are constrained to belong to the friction cone
$\mathcal{F}=\{[\vf{x}_t, x_n] \in\mathbb{R}^3 \,|\, \Vert\vf{x}_t\Vert\le \mu
x_n\}$.

The optimality conditions for Eq. (\ref{eq:maximum_dissipation_principle}) are
\cite{bib:stewart2000rigid, bib:tasora2011}
\begin{flalign}
    &\mu f_{n}\vf{v}_{t} + \lambda \vf{f}_{t} = \vf{0}\nonumber\\
    &0\le \lambda \perp \mu f_{n}-\Vert\vf{f}_t\Vert \ge 0
    \label{eq:mpd_optimality_conditions}
\end{flalign}
where $\lambda$ is the multiplier needed to enforce Coulomb's law condition
$\Vert\vf{f}_t\Vert \le \mu f_{n}$. Notice that in the form we
wrote Eq. (\ref{eq:mpd_optimality_conditions}), multiplier $\lambda$
has units of velocity and it is zero during stiction and takes the value
$\lambda=\Vert\vf{v}_{t}\Vert$ during sliding. Finally, the total contact
force $\vf{f}_c\in\mathbb{R}^3$ expressed in the contact frame $C$ is given by
$\vf{f}_c=[\vf{f}_t\,f_{n}]$.

\subsection{Discrete Time Stepping}

We discretize time into intervals of fixed size
$\delta t$ and seek to advance the state of the system from time $t^n$ to the
next step at $t^{n+1} = t^n + \delta t$. To simplify notation, we use the naught subscript to denote quantities
evaluated at the previous time step $t^n$ while no additional subscript is used
for quantities at the next time step $t^{n+1}$. The full contact problem
consists of the balance of momentum discretized in time together with the full
set of contact constraints, where the unknowns are the next time step
generalized velocities $\mf{v}\in\mathbb{R}^{n_v}$, forces
$\mf{f}\in\mathbb{R}^{3n_c}$ and multipliers ${\bm\lambda}\in\mathbb{R}^{n_c}$
\begin{flalign}
	&\mf{M}_0(\mf{v}-\mf{v}_0) = \delta t\,\mf{k}_0 +
	\delta t\,\mf{J}_0^T\mf{\mf{f}}, 
    \label{eq:scheme_momentum}\\
    &0 \le \phi_{0,i} + (\delta t+d_i\,c_i)\,v_{n,i} + c_i f_{n,i}\nonumber\\
    &\qquad\perp f_{n,i} \ge 0, \quad\qquad\qquad\qquad\! i\in\mathscr{C}(\mf{q}_0)
    \label{eq:scheme_nonpenetration}\\
    &\mu_if_{n,i}\vf{v}_{t,i} + \lambda_i \mf{f}_{t,i} = \vf{0},
    \qquad\qquad\,\, i\in\mathscr{C}(\mf{q}_0)
    \label{eq:scheme_mdp_multiplier}\\
    &0\le \lambda_i \perp \mu_i f_{n,i}-\Vert\mf{f}_{t,i}\Vert \ge 0
    , \!\!\qquad i\in\mathscr{C}(\mf{q}_0)
    \label{eq:scheme_mdp_cone}\\
    &\mf{q} = \mf{q}_0 + \delta t \mf{N}_0\mf{v},
    \label{eq:scheme_q_update}
\end{flalign}
where $\mf{M}_0\in\mathbb{R}^{n_v\times n_v}$ is the mass matrix and
$\mf{k}_0\in\mathbb{R}^{n_v}$ models external forces such as
gravity, gyroscopic terms and other smooth generalized forces such as those
arising from springs and dampers.

We note that typically these velocity-level formulations are written in terms of
impulses $\delta t \mf{\mf{f}}$. The full problem
\eqref{eq:scheme_momentum}-\eqref{eq:scheme_q_update} constitutes a nonlinear
complementarity problem (NCP). \reviewquestion{R3-Q15/R3-Q17/R3-Q21}{Many
variants of this formulation exist in the literature.
\cite{bib:macklin2020primal} introduces both primal and dual formulations of the
problem, \cite{bib:li2020incremental} uses barrier functions along a lagged
dissipative potential to include friction, \cite{bib:stewart2000rigid} uses a
polyhedral approximation of the friction cone to write a linear complementarity
problem (LCP).}

In the next section we describe an approximation that allows one to incorporate the
hydroelastic contact model into velocity-level NCP formulations of this type.
The approach is general in that it can be incorporated into any velocity-level
solver that supports the modeling of compliant point contact.

\section{Overview of the Hydroelastic Contact Model}
\label{sec:hydro_overview}

The hydroelastic contact model \cite{bib:elandt2019pressure} combines two ideas:
\emph{elastic foundation} and \emph{hydrostatic pressure}. Thus the model
introduces an object-centric \emph{virtual} or \emph{elastic} pressure field
$p_e$ to mimic the hydrostatic pressure field of a fluid.
\reviewquestion{R2-Q9/R3-22}{In practice, Drake generates a pressure field for
primitive shapes that is maximum at the medial axis, zero at the boundary, and
linearly interpolated in between. In Drake, users specify how stiff a compliant
object is through the \emph{hydroelastic modulus} $E_h$
\cite{bib:hydroelastics_users_guide}, the value of the pressure field at the
medial axis. How to generate pressure fields for arbitrary non-convex
geometries is currently a topic of active research.}

\begin{figure}[b]
	\centering
	\includesvg[width=0.22\textwidth]{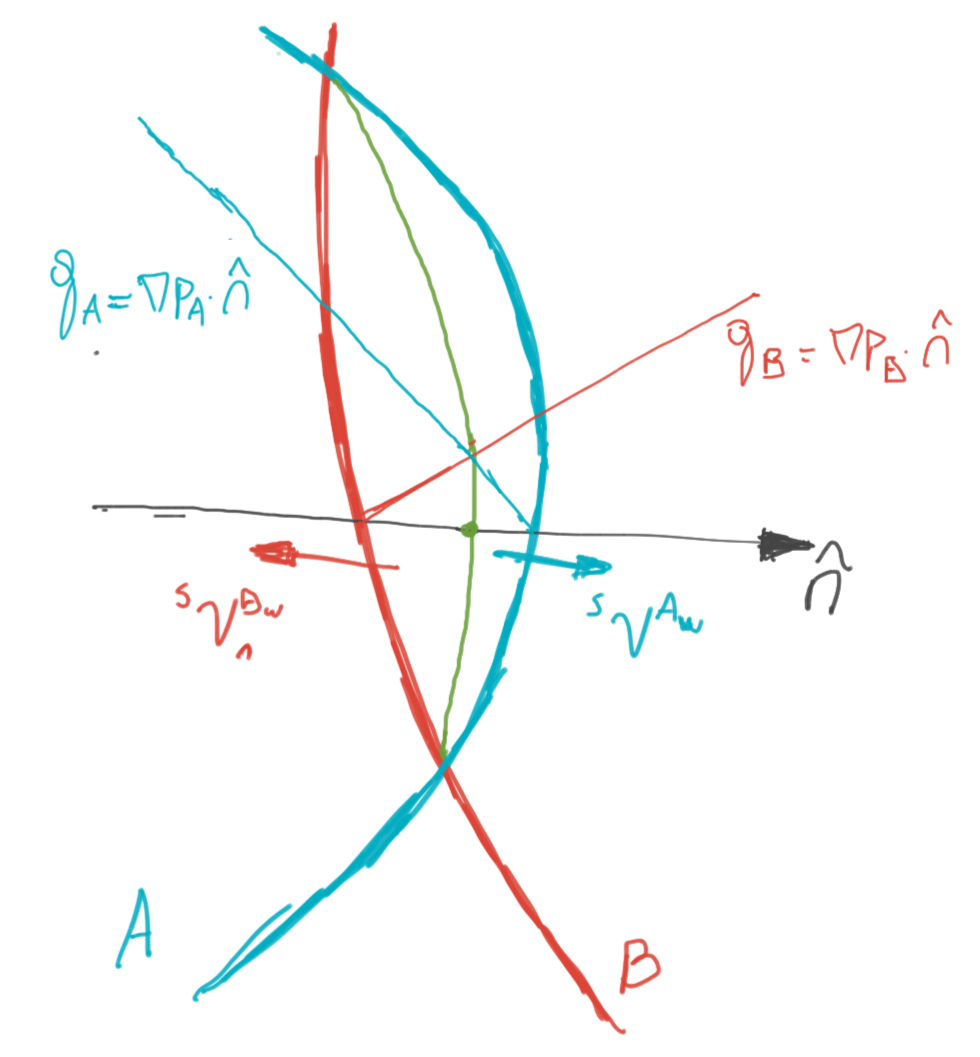}
	\caption{\label{fig:three_dimensional_hydro} 
    Two overlapping compliant bodies with pressure fields $p_{A}$ and $p_{B}$,
    with profiles along the normal direction sketched in dashed lines. The
    contact surface $\mathcal{S}^\cap$ is modeled as the surface of equal
    pressure. We consider the motion of the surface along the normal direction
    $\hat{\vf{n}}$, taking into account directional gradients $g_A =
    -dp_A/dn=-\nabla p_A\cdot\hat{\vf{n}}$ and $g_B = dp_B/dn=\nabla
    p_B\cdot\hat{\vf{n}}$.}
\end{figure}

\reviewquestion{R2-Q7/R3-Q8/R3-26}{Unlike Finite Element models, the
hydroelastic contact model is stateless and the \emph{deformed} configuration of
a body is approximated. Given two overlapping (undeformed) objects $A$ and $B$
with pressure fields $p_{A}$ and $p_{B}$, respectively, the contact surface
$\mathcal{S}^\cap$ is modeled as the surface of equal pressure, see Fig.
\ref{fig:three_dimensional_hydro}}. Total forces and moments on these bodies are
the result of the integral of the equilibrium pressure field $p_e = p_{A} =
p_{B}$ on the contact surface $\mathcal{S}^\cap$.

\subsection{Contact Surface Computation}
\label{sec:contact_surfaces}

We represent the geometry of a compliant body with a tetrahedral volume mesh.
Each vertex of this mesh stores a single scalar pressure value resulting in a
piece-wise linear pressure field $p_e$ which can be used to interpolate pressure
values at any point inside the volume.

\begin{figure}[hb]
  \centering
  \includegraphics[width=0.475\textwidth]{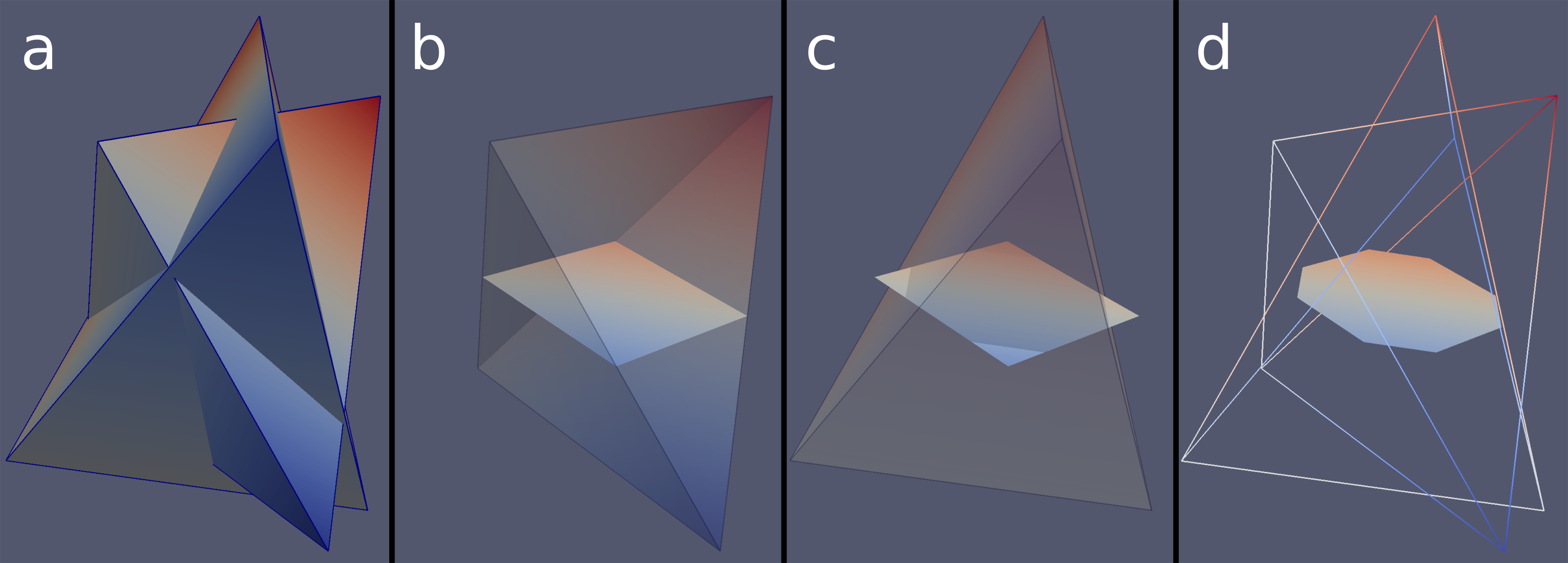}
  \caption{Steps to compute a contact polygon for compliant-compliant contact.
  	       a) Two overlapping tetrahedra. b) Their equilibrium plane is clipped
  	       by the bottom tetrahedron into a square. c) The top tetrahedron clips
  	       the plane further into the final polygon, in this example, an
  	       octagon. d) Contact polygon with linearly interpolated equilibrium
  	       pressure.}
  \label{CompliantTetrahedra}
\end{figure}

The contact surface between two compliant bodies $A$ and $B$ consists of a
number of polygons. We denote with $L_a:\mathbb{R}^3\rightarrow\mathbb{R}$ and
$L_b:\mathbb{R}^3\rightarrow\mathbb{R}$ the linear interpolation of the
respective pressure fields within two tetrahedra $\tau_a \in A$ and $\tau_b \in
B$ having a non-empty intersection. Intersecting tetrahedra can be found
efficiently with a judicious choice of data structures
\cite{bib:elandt2019pressure}. The surface on which $L_a$ equals $L_b$ defines
an equilibrium plane $P_{ab}$. The contact surface is the intersection $P_{ab}
\cap \tau_a \cap \tau_b$, a convex polygon with at most eight vertices, Fig.
\ref{CompliantTetrahedra}. \reviewquestion{R2-Q7}{Recall that \emph{undeformed}
bodies are allowed to overlap, see Fig. \ref{fig:three_dimensional_hydro}.
Therefore intersecting tetrahedra from two bodies as depicted in Fig.
\ref{CompliantTetrahedra} is commonplace within the overlap region of Fig.
\ref{fig:three_dimensional_hydro}.}

While a rigid object can be approximated as a compliant hydroelastic object with
a very large modulus of elasticity, this approach can lead to numerical issues.
Therefore, in Drake, we represent a rigid object solely with a surface mesh of
triangles that tessellates its boundary. In this case, the contact surface
corresponds to the surface of the rigid object clipped by the volume of the
compliant object and the contact pressure $p_e$ is the linear interpolation of
the compliant pressure field onto the contact surface. 

\subsection{Triangulated vs. Polygonal Contact Surfaces}
\label{sec:polygonal_contact_surfaces}

\begin{figure}[hpb]
	\centering
	\includegraphics[width=0.45\textwidth]{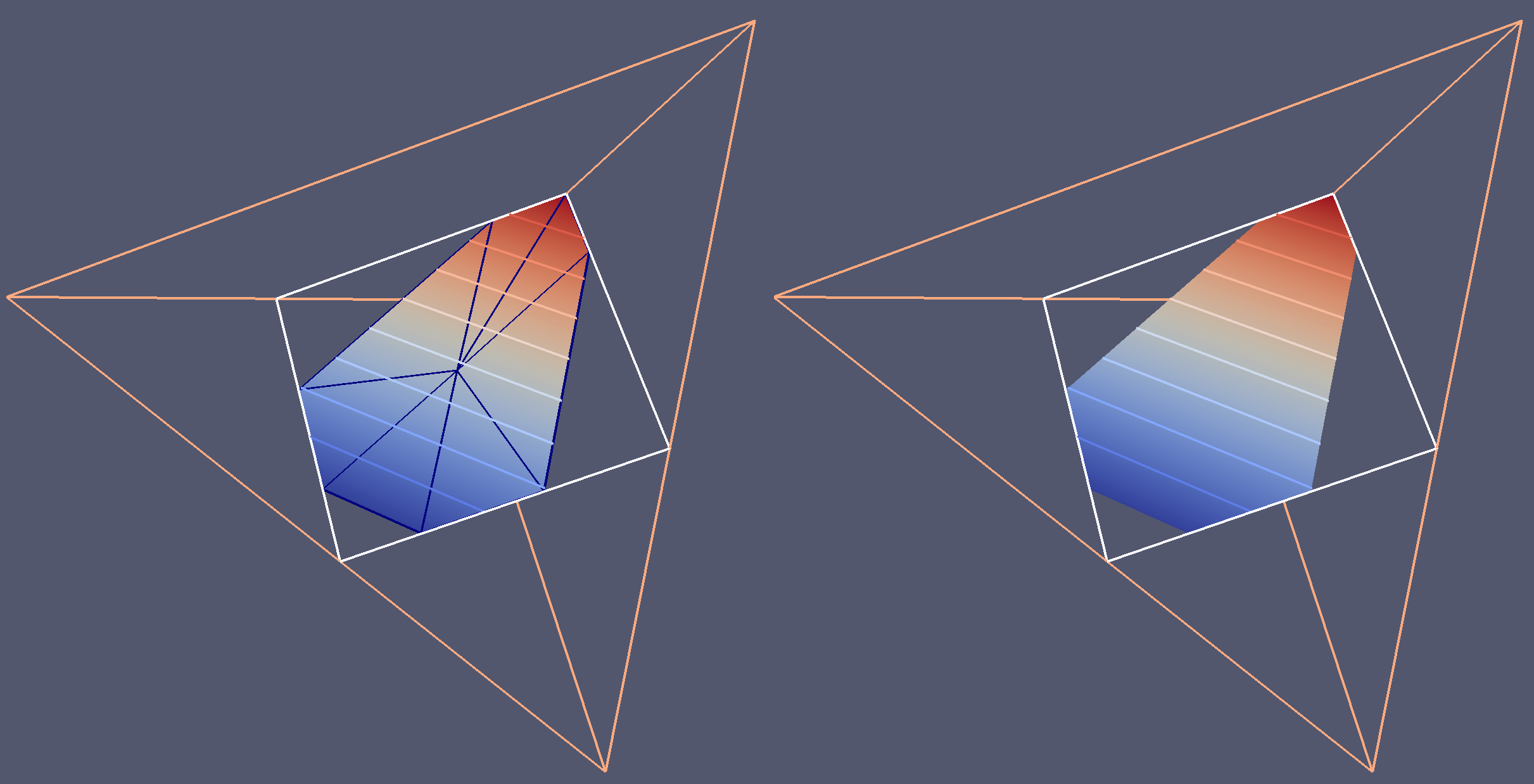}
	\caption{Triangular tessellation (left) as required in
	\cite{bib:elandt2019pressure} and the proposed polygonal approximation
	(right).    
	         The pressure field is linear on the polygon as shown by the contour
             lines and the color shading. The white rectangle outline is a
             visual cue for the spanning plane of the contact polygon relative
             to the compliant tetrahedron drawn in orange outline. }
	\label{TriangulateOrPolygon}
\end{figure}

In \cite{bib:elandt2019pressure} an $n$ sided polygon is divided into a fan of
$n$ triangles that share a vertex at the polygon's centroid, left in Fig.
\ref{TriangulateOrPolygon}. This ensures that only zero area triangles are
added/removed to the contact surface as objects move so that topological
changes do not introduce discontinuities in the contact forces, as required for
error-controlled integration.

As we'll see in Section
\ref{sec:point_contact_approximation}, each face in the contact surface
corresponds to one contact constraint in Eq. \ref{eq:scheme_nonpenetration}.
Therefore to arrive to a smaller contact problem, we seek to minimize the number
of contact constraints and consequently the number of discrete faces. We then
propose to replace the fan of triangles by the original polygon, right in Fig.
\ref{TriangulateOrPolygon}. 

The polygonal representation leads to a significant reduction in the number of
face elements representing the surface, a factor of seven in Fig.
\ref{TriangulateOrPolygon}. Even though much coarser, the polygonal
representation still provides rich contact information and allows to capture
complex area-dependent phenomena. This is demonstrated with test cases of
practical relevance in Section \ref{sec:results_and_discussion}. Finally, the
equilibrium pressure field is linear since it results from the intersection of
the linear pressure fields of overlapping tetrahedra.

\section{Point Contact Approximation}
\label{sec:point_contact_approximation}

\reviewquestion{R1-Q3/R3-Q25}{The key idea introduced in this work is to
approximate the force contribution from each of the polygons described in
Section \ref{sec:polygonal_contact_surfaces} using a first order expansion in
time that resembles the point contact model in Eq.
\eqref{eq:scheme_nonpenetration}. The elastic force contribution $\vf{f}_{e}$
from a polygon with area $A$ is the integral of the pressure field
\begin{equation}
    \vf{f}_{e} = \int_{A} p_e(\vf{x})\,\hat{\vf{n}}\,dA = 
    \hat{\vf{n}}\int_{A} p_e(\vf{x})\,dA
    \label{eq:normal_force_integral}
\end{equation}
where the second equality results from the fact that faces are planar. Moreover,
since the pressure field is linear (Fig. \ref{TriangulateOrPolygon}), this
integral can be computed exactly as
\begin{eqnarray}
    \vf{f}_{e} &=& f_{n,e}\,\hat{\vf{n}},\nonumber\\
    f_{n,e} &=& A\,p_c
    \label{eq:normal_force_analytical}
\end{eqnarray}
where $p_c=p_e(\vf{x}_c)$ is the pressure evaluated at the centroid $\vf{x}_c$
of the polygonal face.}

To obtain an approximation consistent with the discrete framework
\eqref{eq:scheme_momentum}-\eqref{eq:scheme_q_update}, we use a first order
Taylor expansion to approximate the pressure as
\begin{equation}
    p_{c} = \left(p_{c,0} + \delta t\,\frac{dp_c}{dt}\right)_+,
    \label{eq:pressure_expansion}
\end{equation}
where $p_{c,0}$ is the hydroelastic pressure at the previous time step.
\reviewquestion{R3-Q27}{Since pressure is zero at the boundary of each object
and zero outside, we must take the positive part in
\eqref{eq:pressure_expansion} to properly represent this functional form when
bodies break contact.}

We will show next that the time rate of the pressure at the surface can be
approximated as 
\begin{equation}
    \frac{dp_c}{dt} = -g\,v_n,
    \label{eq:pressure_rate_approximation}
\end{equation}
where $g$ is an effective pressure gradient, with units of $\text{Pa}/\text{m}$
and $v_n$ is the normal relative velocity at the centroid. Using this
approximation in Eqs. \eqref{eq:normal_force_analytical} and
\eqref{eq:pressure_expansion}, we can write
\begin{equation}
    f_{n,e} = (-k\phi)_+,
    \label{eq:elastic_force}
\end{equation}
with
\begin{eqnarray}
    k &=& g\,A_0,\nonumber\\
    \phi_0 &=& -\frac{p_{c,0}}{g},\nonumber\\
    \phi &=& \phi_0 + \delta t\,v_n.
    \label{eq:surrogate_quantities}
\end{eqnarray}
where we \emph{froze} geometric quantities at the previous time step. This is
common practice in many discrete time stepping strategies in the literature, see
for instance \cite{bib:stewart1996implicit, bib:anitescu1997formulating}.
Dissipation in \eqref{eq:elastic_force} is incorporated as in
\eqref{eq:linear_normal_force} to obtain an equivalent point contact model.

Using this surrogate signed distance $\phi_0$ and stiffness $k$ we introduce the
contribution of the $i\text{-th}$ face of the contact surface as a compliant
point contact constraint in \eqref{eq:scheme_nonpenetration}.
\reviewquestion{R3-Q5}{Note that the resulting scheme is implicit in the next
time step velocities through \eqref{eq:scheme_nonpenetration}, making the scheme
robust to the choice of time-step size even for stiff materials.}

\reviewquestion{R2-Q4}{Since these quantities are a function of polygon area and
effective pressure gradient, the approximation converges to the original
continuous model in the limit to zero time step. Notice this would not be true
for a simple model where spring-dampers are located at the polygons' centroids.}

\subsection{Pressure Time Rate}

At a given point on the contact surface in Fig.
\ref{fig:three_dimensional_hydro} we analyze the relative motion of bodies $A$
and $B$ in the direction normal to the surface. We define a coordinate $x$ in
the normal direction such that $x=0$ at the surface and it increases in the
direction along the normal.

Along this normal direction, in the neighborhood to the contact point, we
approximate pressure fields $p_A(x)$ and $p_B(x)$ as linear functions of the
coordinate $x$
\begin{align}
    p_A(x) &= -g_A(x-x_A(t)) + b_{A},\\
    p_B(x) &=  g_B(x-x_B(t)) + b_{B},
\end{align}
where $g_A=-\nabla p_A\cdot\hat{\vf{n}}$ and $g_B=\nabla p_B\cdot\hat{\vf{n}}$
are the slopes along the normal, $x_A(t)$ and $x_B(t)$ are points rigidly
affixed to $A$ and $B$, respectively, and $b_{A}$ and $b_{B}$ are simply the
pressure values at $x_A(t)$ and $x_B(t)$, respectively. This is a reasonable
approximation given that the pressure fields are piecewise linear functions
within each compliant volume.

The equilibrium pressure at the surface, $x=0$, is found by equating the
hydroelastic pressures
\begin{eqnarray}
    p_e = g_A\,x_A(t) + b_{A}= -g_B\,x_B(t) + b_{B}.
    \label{eq:one_dimenional_pressure_balance}
\end{eqnarray}

We take the time derivative of (\ref{eq:one_dimenional_pressure_balance}) to
find the rate of change of the pressure, as we need it in
\eqref{eq:pressure_expansion} at each polygon centroid
\begin{equation}
    \frac{d p_e}{dt} = g_A\,v_A(t) = -g_B\,v_B(t),
    \label{eq:one_dimenional_velocity_balance}
\end{equation}
where $v_A(t)$ and $v_B(t)$ are the respective velocities of each body along the
normal. These velocities are \emph{relative to the contact surface} since
coordinate $x$ is defined relative to the surface, located at all times at
$x=0$. Since the pressure fields are fixed in the body frames,
$\dot{b}_{A}=\dot{b}_{B}=0$. In terms of these velocities, the normal velocity
$v_n$ is given by
\begin{eqnarray}
    v_n = v_B - v_A
    \label{eq:normal_relative_velocity}
\end{eqnarray}

Combining Eqs. \eqref{eq:one_dimenional_velocity_balance} and
\eqref{eq:normal_relative_velocity} we can write velocities $v_A(t)$ and
$v_B(t)$ in terms of the normal velocity $v_n$ as
\begin{eqnarray}
    v_A &=& -\frac{g_B}{g_A+g_B}\,v_n,\nonumber\\
    v_B &=& \frac{g_A}{g_A+g_B}\,v_n.
    \label{eq:relative_velocities}
\end{eqnarray}

The final expression for the rate of change of the pressure at the interface is
obtained using the relative velocities from \eqref{eq:relative_velocities} into
\eqref{eq:one_dimenional_velocity_balance}. After some minimal algebraic
manipulation, the result is
\begin{equation}
    \frac{d p_e}{dt} = \frac{-g_A\,g_B}{g_A+g_B}v_n.
    \label{eq:one_dimensional_pressure_rate}
\end{equation}

Typically the pressure gradients and the normal direction align
along the same line and therefore both $g_A$ and $g_B$ are positive. In this
case $dp/dt < 0$ for $v_n>0$ and the pressure decreases as the bodies move away
from each other, as expected. However, special care must be taken when $g_A < 0$
or $g_B < 0$.  Since the discrete approximation of point contact requires $k>0$,
we simply ignore polygons where the conditions $g_A > 0$ and $g_B > 0$ are not
satisfied. We find that this is not a major problem in practice since this
situation corresponds to corner cases of the hydroelastic contact model for
which \emph{pushing into the object} leads to a decrease of the contact forces
instead of an increase as expected.

\section{Results and Discussion}
\label{sec:results_and_discussion}

We present a series of simulation cases to assess the robustness, accuracy, and
performance of our method. The time step for each simulation is chosen such
that it can properly resolve the dynamics of each specific problem. It is a
trade off between accuracy and speed.

In Drake we have two velocity level solvers; TAMSI
\cite{bib:castro2020transition} and SAP \cite{bib:castro2021sap}. SAP uses a
convex approximation of contact excellent for problems dominated by stiction or
sliding at low velocities. We use SAP in Section \ref{sec:pancake_flip} for our
scalability studies since it uses supernodal sparse algebra and TAMSI everywhere
else.

\subsection{Sliding and Spinning Disk}
\label{sec:disk_spin}
To assess the accuracy of our method's ability to capture the highly non-linear
coupling between net force and torque, we study a sliding and spinning disk with
a known analytical solution \cite{bib:farkas2003frictional}.

Based on the dimensions of a U.S. quarter dollar coin, we simulate a disk of
radius $R = 1.213\text{ cm}$, thickness $t = 1.75\text{ mm}$, mass $m =
5.67\text{ g}$, friction coefficient $\mu = 0.2$, and hydroelastic modulus $E_h
= 1.0 \text{ GPa}$ lying flat on a horizontal plane set into motion with initial
values of translational velocity $v$ and angular velocity $\omega$. The
analytical result for this example establishes a dimensionless parameter
$\varepsilon = \frac{v}{\omega R}$ that, regardless of initial conditions,
converges to $\varepsilon^* \approx 0.653$ as the coin comes to rest. We set
initial angular and translational velocities to span initial values
$\varepsilon_0$ in the range $[0.1; 10]$.

A fan of 152 triangles discretizes the circular geometry of the coin. To
estimate the error introduced by the discrete geometry, we first simulate our
model using error controlled integration to a tight accuracy of $10^{-2}\%$. We
find the numerical solution with discrete geometry converges to
$\varepsilon_\text{disc}^* = 0.64426$, at only $1.3\%$ error from
$\varepsilon^*$.

\begin{figure}[!h]
	\centering
	\includegraphics[width=0.4\textwidth]{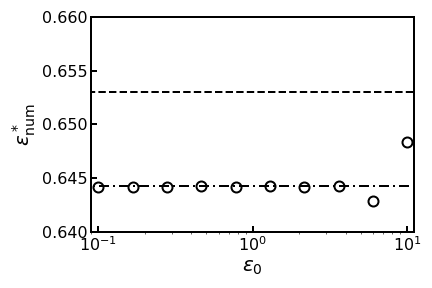}
	\caption{\label{fig:spinning_coin_epsilon_num}
    Numerically computed $\varepsilon_\text{num}^*$ vs. initial $\varepsilon_0$
	(circles). Values are within $0.01\text{-}0.5 \%$ of the numerical reference
	$\varepsilon_\text{disc}^* = 0.64426$ (dash-dotted line) and within $1.3\%$
	of the theoretical value $\varepsilon^* = 0.653$ (dashed line).}
\end{figure}

We now use our velocity-level discrete solver with a fixed time step of $\delta
t = 10^{-3}\text{ s}$ to compute numerical approximations
$\varepsilon_\text{num}^*$ from various initial conditions. Theory
\cite{bib:farkas2003frictional} predicts a constant $\varepsilon^*$ regardless
of the initial conditions. The numerical results confirm this prediction within
$0.01\text{-}0.5 \%$ of $\varepsilon_\text{disc}^*$ and within $1.3\%$ of
$\varepsilon^*$, see Fig. \ref{fig:spinning_coin_epsilon_num}.
\reviewquestion{R3-Q32}{Variations in these results are caused by numerical
sensitivity to the zero-over-zero limit in $\varepsilon = v/(\omega R)$ as the
disk comes to rest.}

\begin{figure}[h!]
	\centering
	\includegraphics[trim={0.5cm 0 0.4cm 0},clip, width=0.23\textwidth]{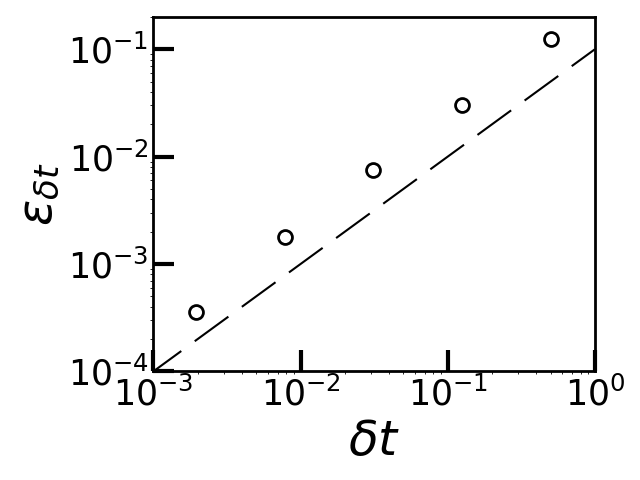}
    \includegraphics[trim={0.5cm 0 0.4cm 0},clip, width=0.23\textwidth]{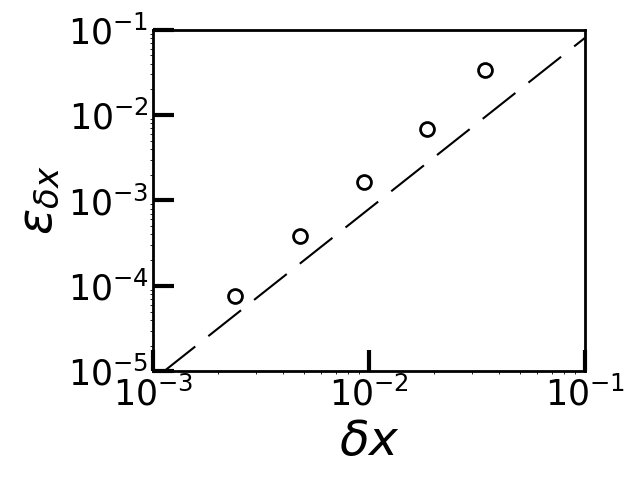}
	\caption{\label{fig:spinning_coin_convergence} Convergence study with time
	step size (left, $\delta x = 2.4\text{ mm}$) and with grid size (right,
	$\delta t = 1.0\text{ ms}$). Dashed references lines are shown for first
	order on the left and for second order on the right.}
\end{figure}

Finally, we perform a convergence study to verify the convergence of our method.
For the reference solutions used in Fig. \ref{fig:spinning_coin_convergence} we
use a time step size and grid size an order of magnitude smaller than the
smallest size shown in the figures. For each timestep $\delta t$ we define the
relative error of the computed trajectory $x_{\delta t}(t)$ vs the reference
trajector $x(t)$ as:
\[ \varepsilon_{\delta t} = || x(t)_{\delta t} - x(t)||_2 \big/ ||x(t)||_2\]
and likewise for $\varepsilon_{\delta x}$. Our solver TAMSI
\cite{bib:castro2020transition} is first order accurate in time, which is
verified with the time step convergence study in Fig.
\ref{fig:spinning_coin_convergence}. Even though we show that a single point at
the centroid of each polygon integrates pressure exactly, Fig.
\ref{fig:spinning_coin_convergence} shows a quadratic convergence with grid
size. This is due to the fact that moments are proportional to both pressure and
position, and therefore are not integrated exactly but with a truncation error
quadratic on the grid size. If case this is not clear, the \emph{integration} of
moments is being accounted for by the term $\mf{J}_0^T\mf{\mf{f}}$ in Eq.
\eqref{eq:scheme_momentum}, which effectively accumulates the contributions from
each polygon onto the corresponding body.

\subsection{Pancake Flip}
\label{sec:pancake_flip}
In this scenario, a Kinova JACO arm (6 DOF) is outfitted with a highly compliant
\textit{Soft-bubble} gripper \cite{bib:kuppuswamy2020soft}. The arm is anchored
to a table which has a stand holding a spatula, a cylindrical stove top, and a
pancake, modeled as a flat ellipsoid, on top. The \textit{Soft-bubble} gripper
and the pancake are modeled as compliant objects with hydroelastic modulus $E_h
= 10^{5}\text{ Pa}$ and $E_h = 10^{4}\text{ Pa}$, respectively.
\reviewquestion{R3-Q8}{Even though pancakes fold in reality, synthetic silicone
pancakes were used in the real experimental setup, and therefore hydroelastic
contact proved to be a useuful approximation.} All remaining objects are modeled
as rigid.

\begin{figure}[b]
  \centering
  \includegraphics[width=0.41\textwidth]{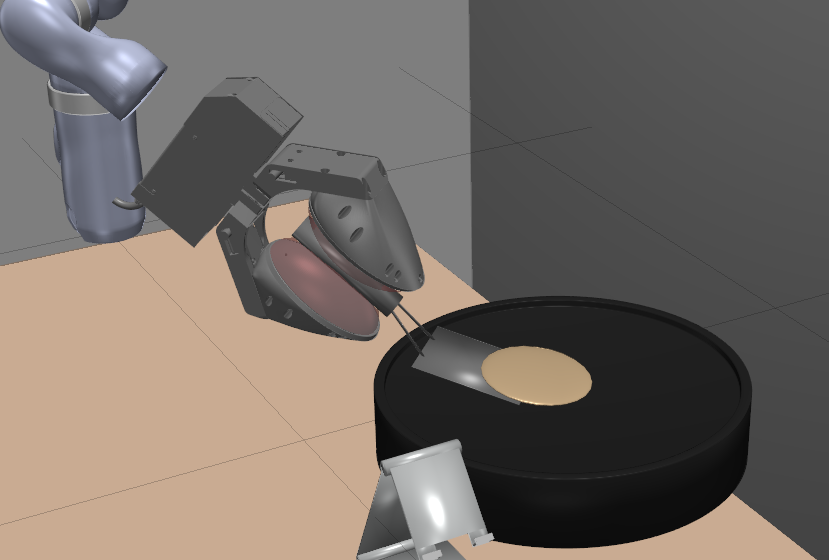}
  \caption{\label{fig:pancake_flip} The scoop process of the pancake flip task.
  See associated video.}
\end{figure}

The controller process tracks a prescribed sequence of Cartesian end-effector
keyframe poses. We use force feedback to gauge successful grasps and to know
when the spatula makes contact with the stove top. 

The robot is commanded to grab the spatula from the stand and subsequently
scoop, raise, and flip the pancake over on the stove, see Fig.
\ref{fig:pancake_flip} and the accompanying supplemental video.

\begin{figure}[h]
  \centering
  \includegraphics[width=0.48\textwidth]{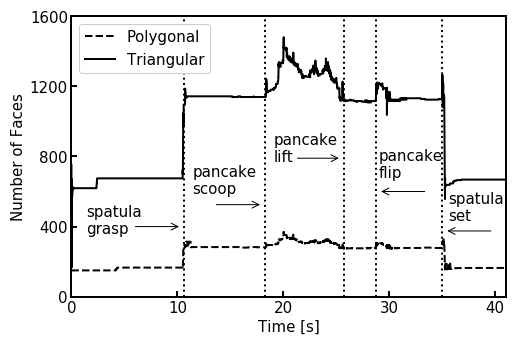}
  \caption{\label{fig:number_of_features} Number of faces generated as a
  function of time. Important events during the task are highlighted.}
\end{figure}

Figure \ref{fig:number_of_features} shows the number of faces throughout the
simulation using both triangular and polygonal tessellations. On average, the
number of faces is 4.05 times smaller when using the polygonal tessellation.
Still, the model is able to resolve the net torque on the spatula needed to
achieve a secure grasp. Moreover, with the resulting reduction in the number of
contact constraints, our solver performs 4.09 times faster. The computation of
polygonal tessellations is only about 10\% faster than the corresponding
triangular tessellations.

\begin{figure}[h]
  \centering
  \includegraphics[width=0.35\textwidth]{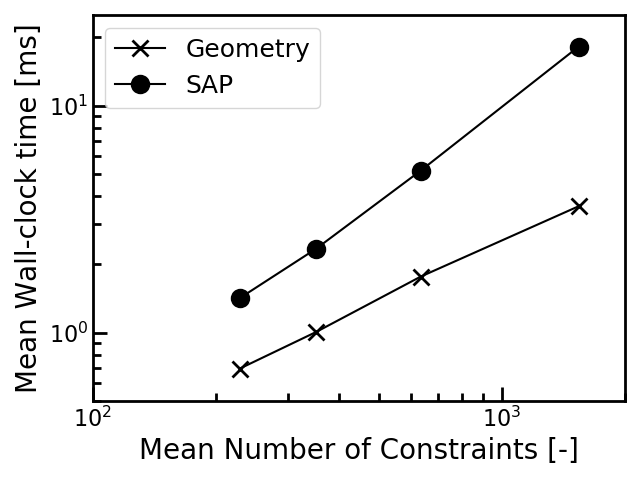}
  \caption{\label{fig:pancake_flip_scalability} Mean wall-clock per time step
  vs. mean number of constraints per time step, for geometry and SAP solver.}
\end{figure}

\reviewquestion{R1-Q5}{To assess scalability and task success at different grid
sizes, we performed a grid refinement study using Drake's SAP solver
\cite{bib:castro2021sap}}. We used a system with 24 2.2 GHz Intel Xeon cores
(E5-2650 v4) and 128 GB of RAM, running Linux. However we run in a single
thread. We use the steady clock from the STL \verb+std::chrono+ library to
measure wall-clock time. All grids use polygonal tessellations. Our coarsest
grids result in 230 contacts per time step on average and we progressively
refine grids by a factor of two, resulting in about 1500 constraints per time
step on average, see the accompanying video. Figure
\ref{fig:pancake_flip_scalability} shows wall-clock time for the geometric
queries and for the solver as a function of the average number of constraints
per time-step.

We observe that the cost of the geometric queries is linear with the number of
constraints (faces), demonstrating the effectiveness of OBBs as acceleration
data structures in our implementation. For fully dense problems, we expect the
solver to have $\mathcal{O}(n^3)$ complexity, where $n$ denotes the number of
variables. For sparse problems, the complexity is $\mathcal{O}(d^3)$, where $d$
is the size of the largest clique in a chordal completion of the linear system
matrix. A best fit exponent is about $\sim 1.3$ in Fig.
\ref{fig:pancake_flip_scalability}, demonstrating the effectiveness of the
supernodal algebra, even though this case is not very sparse.

For our coarsest set of grids, the spatula resembles a box rather than the
original cylinder shape and the gripper uses a similarly coarse grid. Still, the
robot completed the task successfully at all grid refinement levels. This
demonstrates that the completion of this task in simulation is rather
insensitive to mesh resolution.

Finally, our colleagues at TRI prototyped controllers in simulation that
transferred seamlessly to the real robotic system, as shown in the accompanying
video.

\subsection{Spatula Slip Control}
\begin{figure}[h!]
	\centering
	\includegraphics[width=0.23\textwidth]{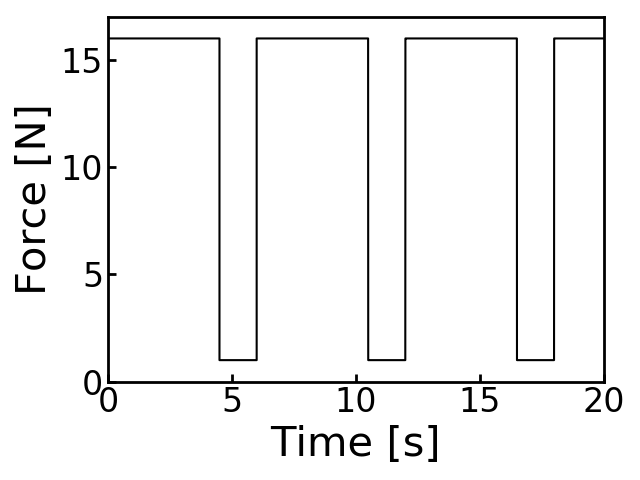}
    \includegraphics[width=0.23\textwidth]{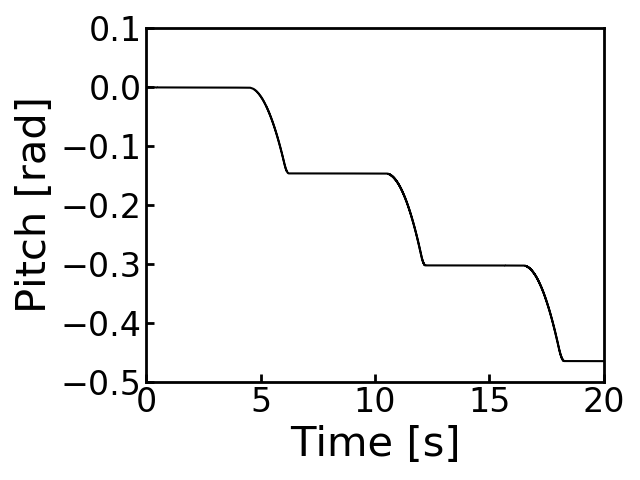}
	\caption{\label{fig:spatula_pitch} Commanded grasp force (left) and spatula
	pitch (right).}
\end{figure}

We now demonstrate the effectiveness of our method to capture area-dependent
phenomena such as the net torque required to successfully grasp an object. We
simulate the aforementioned \textit{Soft-bubble} gripper
\cite{bib:kuppuswamy2020soft} anchored to the world holding a spatula by the
handle horizontally. The grasp force is commanded to vary between $1\text{ N}$
and $16\text{ N}$ with square wave having a 6 second period and a 75\% duty
cycle, left on Fig. \ref{fig:spatula_pitch}. This controller results in a
periodic transition from a secure grasp with stiction to a loose grasp where the
spatula is allowed to rotate within the grasp in a controlled manner, see Fig.
\ref{fig:slip_control} and the accompanying video.

Figure \ref{fig:slip_control} (left) shows a closeup of the contact geometry
used for this model. Notice that while well resolved, we use a rather coarse
tessellation of the compliant bubble surfaces of the gripper. The polygonal
tessellation provides a rich representation of the contact patch exhibited by an
elongated shape induced by the geometry of the handle, Fig.
\ref{fig:slip_control} (right).

This level of grasp control is achieved by properly resolving contact patch area
changes; this degree of control would be very difficult, if not impossible, to
emulate using point contact approaches. 

\section{Limitations}
\label{sec:limitations}

All models are approximations of reality. We would like to explicitly state
the limitations of our approach:
\begin{itemize}
    \item \reviewquestion{R3-Q4}{Acceleration data structures and linear
    tetrahedra are key for a performant implementation of the hydroelastic
    contact model \cite{bib:elandt2019pressure} for simulations at real time
    rates. Thus far, this limits the implementation to linear elements. Higher
    order elements or alternative representations are in interesting research
    direction.}
    \item We show in Section \ref{sec:point_contact_approximation} that a single
    quadrature point located at the centroid of a polygon integrates the
    linear pressure field exactly. Conversely, moments are not integrated
    exactly. However, our method achieves second order accuracy with grid size
    as demonstrated with the grid study in Section \ref{sec:disk_spin}.
    \item \reviewquestion{R3-Q8}{The hydroelastic contact model is a modeling
    approximation which does not introduce deformation state. Therefore the
    model cannot resolve large deformations phenomena such as buckling or
    folding.
    \item Thin objects can be problematic. For instance, an equilibrium surface
    could not exit if a compliant thin box is pushed deep enough into a
    compliant half space. For this to happen the thin box needs to first get
    into such a configuration, which is possible especially for large time
    steps. We are currently working on ways to remedy this problem.}
\end{itemize}

\section{Conclusions}
\label{sec:conclusions}

We presented a discrete in time approximation of the hydroelastic contact model to
enable simulation of contact rich patches using velocity-level discrete solvers
for simulation at real-time rates. The approach is general enough in that it can
be incorporated into any velocity level solver that can handle compliant point
contact.

We demonstrated the highly predictive nature of this model in a test case with
strong coupling between net force and torque, matching known analytical results
to within 1.3\% without parameter tuning beyond choosing a mesh that can
reasonably represent the geometry and choosing a time step that can resolve the
temporal dynamics of the problem. Even though the polygonal tessellations are
coarser than the original triangular tessellations from
\cite{bib:elandt2019pressure}, we demonstrated the effectiveness of the approach
to predict area-dependent phenomena such as the net torque required for the
successful completion of a manipulation task.

Our novel surface representation in terms of polygonal faces leads to a drastic
reduction in the number of contact constraints, a significantly smaller contact
problem at each time step, and consequently a substantial speedup enabling
simulation at interactive rates.

We present both time step size and grid size studies in order to assess the
expected rate of convergence of our approximations. In particular, our method
converges quadratically with grid size. The order of convergence with time step
size depends on the particulars of the velocity level formulation, first order
for our TAMSI \cite{bib:castro2020transition} solver.

Finally, we include a mesh refinement study on the simulation of a real robotic
task that involves grasping. The study reveals that the success of the task is
not very sensitive to mesh resolution, even when using very coarse grids.
Moreover, the study allowed us to assess the scalability of the contact queries
and our SAP solver \cite{bib:castro2021sap} with the number of constraints.

The hydroelastic contact model and the discrete approximation presented in this
work are made available in the open-source robotics toolbox Drake
\cite{bib:drake}. The new model has been used extensively for work conducted at
the Toyota Research Institute on prototyping and validating controllers for
dexterous manipulation of complex geometries \cite{bib:russ2021_blog}.

\addtolength{\textheight}{-12cm}   

\section*{ACKNOWLEDGMENT}

We thank the reviewers for their feedback, Sean Curtis for his invaluable
support on geometry and Drake implementation details, Michael Sherman for his
insight and helpful discussions, and Naveen Kuppuswamy for the Sim-to-Real
pancake demo setup.


\bibliographystyle{IEEEtran}
\bibliography{IEEEabrv,discrete_hydro}

\begin{thebibliography}{10}
\providecommand{\url}[1]{#1}
\csname url@rmstyle\endcsname
\providecommand{\newblock}{\relax}
\providecommand{\bibinfo}[2]{#2}
\providecommand\BIBentrySTDinterwordspacing{\spaceskip=0pt\relax}
\providecommand\BIBentryALTinterwordstretchfactor{4}
\providecommand\BIBentryALTinterwordspacing{\spaceskip=\fontdimen2\font plus
\BIBentryALTinterwordstretchfactor\fontdimen3\font minus
  \fontdimen4\font\relax}
\providecommand\BIBforeignlanguage[2]{{%
\expandafter\ifx\csname l@#1\endcsname\relax
\typeout{** WARNING: IEEEtran.bst: No hyphenation pattern has been}%
\typeout{** loaded for the language `#1'. Using the pattern for}%
\typeout{** the default language instead.}%
\else
\language=\csname l@#1\endcsname
\fi
#2}}

\bibitem{bib:kuppuswamy2020soft}
N.~Kuppuswamy, A.~Alspach, A.~Uttamchandani, S.~Creasey, T.~Ikeda, and
  R.~Tedrake, ``Soft-bubble grippers for robust and perceptive manipulation,''
  in \emph{2020 IEEE/RSJ International Conference on Intelligent Robots and
  Systems (IROS)}.\hskip 1em plus 0.5em minus 0.4em\relax IEEE, 2020, pp.
  9917--9924.

\bibitem{bib:catto_softconstraints}
E.~Catto, ``Soft constraints: Reinventing the spring,'' Game Developer
  Conference, 2011.

\bibitem{bib:luo2006compliant}
L.~Luo and M.~Nahon, ``A compliant contact model including interference
  geometry for polyhedral objects,'' 2006.

\bibitem{bib:gonthier2007contact}
Y.~Gonthier, ``Contact dynamics modelling for robotic task simulation,'' 2007.

\bibitem{bib:wakisaka2017loosely}
N.~Wakisaka and T.~Sugihara, ``Loosely-constrained volumetric contact force
  computation for rigid body simulation,'' in \emph{2017 IEEE/RSJ International
  Conference on Intelligent Robots and Systems (IROS)}.\hskip 1em plus 0.5em
  minus 0.4em\relax IEEE, 2017, pp. 6428--6433.

\bibitem{bib:bullet_siggraph2015}
E.~Coumans, ``{SIGGRAPH 2015} course,''
  \url{https://pybullet.org/wordpress/index.php/forum-2}, 2015.

\bibitem{bib:moravanszky2004fast}
A.~Moravanszky, P.~Terdiman, and A.~Kirmse, ``Fast contact reduction for
  dynamics simulation,'' \emph{Game programming gems}, vol.~4, pp. 253--263,
  2004.

\bibitem{bib:erleben2018methodology}
K.~Erleben, ``Methodology for assessing mesh-based contact point methods,''
  \emph{ACM Transactions on Graphics (TOG)}, vol.~37, no.~3, pp. 1--30, 2018.

\bibitem{bib:johnson1987contact}
K.~L. Johnson, \emph{Contact mechanics}.\hskip 1em plus 0.5em minus 0.4em\relax
  Cambridge University Press, 1987.

\bibitem{bib:sherman2011simbody}
M.~A. Sherman, A.~Seth, and S.~L. Delp, ``Simbody: multibody dynamics for
  biomedical research,'' \emph{Procedia IUTAM}, vol.~2, pp. 241--261, 2011.

\bibitem{bib:elandt2019pressure}
R.~Elandt, E.~Drumwright, M.~Sherman, and A.~Ruina, ``A pressure field model
  for fast, robust approximation of net contact force and moment between
  nominally rigid objects,'' in \emph{2019 IEEE/RSJ International Conference on
  Intelligent Robots and Systems (IROS)}.\hskip 1em plus 0.5em minus
  0.4em\relax IEEE, 2019, pp. 8238--8245.

\bibitem{bib:macklin2020local}
M.~Macklin, K.~Erleben, M.~M{\"u}ller, N.~Chentanez, S.~Jeschke, and Z.~Corse,
  ``Local optimization for robust signed distance field collision,''
  \emph{Proceedings of the ACM on Computer Graphics and Interactive
  Techniques}, vol.~3, no.~1, pp. 1--17, 2020.

\bibitem{bib:tang2012continuous}
M.~Tang, D.~Manocha, M.~A. Otaduy, and R.~Tong, ``Continuous penalty forces,''
  \emph{ACM Transactions on Graphics (TOG)}, vol.~31, no.~4, pp. 1--9, 2012.

\bibitem{bib:geilinger2020add}
M.~Geilinger, D.~Hahn, J.~Zehnder, M.~B{\"a}cher, B.~Thomaszewski, and
  S.~Coros, ``Add: Analytically differentiable dynamics for multi-body systems
  with frictional contact,'' \emph{ACM Transactions on Graphics (TOG)},
  vol.~39, no.~6, pp. 1--15, 2020.

\bibitem{bib:drake}
R.~Tedrake and the Drake Development~Team, ``Drake: Model-based design and
  verification for robotics,'' https://drake.mit.edu, 2019.

\bibitem{bib:ode}
R.~Smith, ``Open dynamics engine,'' \url{http://www.ode.org}.

\bibitem{bib:dart}
J.~Lee, M.~X. Grey, S.~Ha, T.~Kunz, S.~Jain, Y.~Ye, S.~S. Srinivasa,
  M.~Stilman, and C.~K. Liu, ``Dart: Dynamic animation and robotics toolkit,''
  \emph{Journal of Open Source Software}, vol.~3, no.~22, p. 500, 2018.

\bibitem{bib:vortex}
{CM Labs Simulations}, ``Theory guide: Vortex software’s multibody dynamics
  engine,'' \url{https://www.cm-labs.com/vortexstudiodocumentation}.

\bibitem{bib:mujoco}
E.~Todorov, ``{MuJoCo},'' \url{http://www.mujoco.org}.

\bibitem{bib:castro2020transition}
A.~M. Castro, A.~Qu, N.~Kuppuswamy, A.~Alspach, and M.~Sherman, ``A
  transition-aware method for the simulation of compliant contact with
  regularized friction,'' \emph{IEEE Robotics and Automation Letters}, vol.~5,
  no.~2, pp. 1859--1866, 2020.

\bibitem{bib:castro2021sap}
A.~Castro, F.~Permenter, and X.~Han, ``An unconstrained convex formulation of
  compliant contact,'' 2021, preprint available at
  \url{https://arxiv.org/abs/2110.10107}.

\bibitem{bib:lacoursiere2011spook}
C.~Lacoursiere and M.~Linde, ``Spook: a variational time-stepping scheme for
  rigid multibody systems subject to dry frictional contacts,'' \emph{UMINF
  report}, vol.~11, 2011.

\bibitem{bib:stewart2000rigid}
D.~E. Stewart, ``Rigid-body dynamics with friction and impact,'' \emph{SIAM
  review}, vol.~42, no.~1, pp. 3--39, 2000.

\bibitem{bib:tasora2011}
A.~Tasora and M.~Anitescu, ``A matrix-free cone complementarity approach for
  solving large-scale, nonsmooth, rigid body dynamics,'' \emph{Computer Methods
  in Applied Mechanics and Engineering}, vol. 200, no. 5-8, pp. 439--453, 2011.

\bibitem{bib:macklin2020primal}
M.~Macklin, K.~Erleben, M.~M{\"u}ller, N.~Chentanez, S.~Jeschke, and T.-Y. Kim,
  ``Primal/dual descent methods for dynamics,'' in \emph{Computer Graphics
  Forum}, vol.~39, no.~8.\hskip 1em plus 0.5em minus 0.4em\relax Wiley Online
  Library, 2020, pp. 89--100.

\bibitem{bib:li2020incremental}
M.~Li, Z.~Ferguson, T.~Schneider, T.~Langlois, D.~Zorin, D.~Panozzo, C.~Jiang,
  and D.~M. Kaufman, ``Incremental potential contact: Intersection-and
  inversion-free, large-deformation dynamics,'' \emph{ACM transactions on
  graphics}, 2020.

\bibitem{bib:hydroelastics_users_guide}
{Drake Development Team}, ``Hydroelastic contact user guide,''
  \url{https://drake.mit.edu/doxygen_cxx/group__hydroelastic__user__guide.html}.

\bibitem{bib:stewart1996implicit}
D.~E. Stewart and J.~C. Trinkle, ``An implicit time-stepping scheme for rigid
  body dynamics with inelastic collisions and coulomb friction,''
  \emph{International Journal for Numerical Methods in Engineering}, vol.~39,
  no.~15, pp. 2673--2691, 1996.

\bibitem{bib:anitescu1997formulating}
M.~Anitescu and F.~A. Potra, ``Formulating dynamic multi-rigid-body contact
  problems with friction as solvable linear complementarity problems,''
  \emph{Nonlinear Dynamics}, vol.~14, no.~3, pp. 231--247, 1997.

\bibitem{bib:farkas2003frictional}
Z.~Farkas, G.~Bartels, T.~Unger, and D.~E. Wolf, ``Frictional coupling between
  sliding and spinning motion,'' \emph{Physical review letters}, vol.~90,
  no.~24, p. 248302, 2003.

\bibitem{bib:russ2021_blog}
R.~Tedrake, ``Drake: Model-based design in the age of robotics and machine
  learning,'' \url{https://medium.com/toyotaresearch/robotics/home}, May 2021.

\end{thebibliography}

\end{document}